\documentclass[prl,aps,singlecolumn,superscriptaddress,preprintnumbers,nopacs,floatfix,amsmath,amssymb]{revtex4}

\usepackage{graphicx}
\usepackage{dcolumn}
\usepackage{bm}

\usepackage{amsmath}
\usepackage{graphicx}
\usepackage{amssymb}
\usepackage{mathrsfs}
\usepackage{bbm}
\usepackage{epsfig}

\newcommand{\la}{\label}

\newcommand{\be}{\begin{eqnarray}}
\newcommand{\ee}{\end{eqnarray}}

\begin{document}

%
\title{ Long range correlations and folding angle in polymers  \\
with applications to  $\alpha$-helical proteins
}

\author{Andrey Krokhotin}
\email{Andrei.Krokhotine@cern.ch}
\affiliation{Department of Physics and Astronomy, Uppsala University,
P.O. Box 803, S-75108, Uppsala, Sweden}
\author{Stam Nicolis}
\email{Stam.Nicolis@lmpt.univ-tours.fr}
\affiliation{
Laboratoire de Mathematiques et Physique Theorique
CNRS UMR 6083, F\'ed\'eration Denis Poisson, Universit\'e de Tours,
Parc de Grandmont, F37200, Tours, France}
\author{Antti J. Niemi}
\email{Antti.Niemi@physics.uu.se}
\homepage{http://www.folding-protein.org}
\affiliation{Department of Physics and Astronomy, Uppsala University,
P.O. Box 803, S-75108, Uppsala, Sweden}
\affiliation{
Laboratoire de Mathematiques et Physique Theorique
CNRS UMR 6083, F\'ed\'eration Denis Poisson, Universit\'e de Tours,
Parc de Grandmont, F37200, Tours, France}
\affiliation{Department of Physics, Beijing Institute of Technology,
Haidian District, Beijing 100081, P. R. China}

\begin{abstract}
\noindent
The conformational complexity of linear polymers far exceeds that
of point-like atoms and molecules. Polymers  can bend, twist, even become knotted.
Thus they may also display a much richer phase structure than point particles.  
But it is not very easy to characterize the phase of a polymer.  
Essentially,  the only attribute is the radius of gyration. The way how it changes 
when the degree of polymerization becomes different, and how it evolves 
when the ambient temperature and solvent properties change,
discloses the phase of the polymer.  Moreover, in any finite length chain
there are corrections to scaling, that complicate the detailed analysis 
of the phase structure. Here we introduce a quantity that we call the folding 
angle,  a novel tool  to identify and scrutinize the phases of polymers. 
We argue for a mean-field relationship between its values and those of the scaling exponent 
in the radius of gyration. But unlike in the case of the radius of gyration, 
the value of the folding angle can be evaluated  from a single structure.
As an example we estimate the value of the 
folding angle in the case of crystallographic $\alpha$-helical
protein structures in the Protein Data Bank (PDB).
We also show how the value can be numerically computed using a theoretical 
model of $\alpha$-helical chiral homopolymers.
\end{abstract}



\maketitle

Despite substantial differences in their chemical composition, 
all linear polymers are presumed to share the same universal phase 
structure \cite{degennes,schafer,naka}. But the phase where a particular polymer resides 
depends on many factors including polymer concentration, 
the quality of solvent,  ambient
temperature and  pressure.  Three phases are commonly identified, each 
of them categorized by the manner how the polymer fills the space 
\cite{degennes,schafer,naka,naka2}:
If the solvent is poor and the attractive
interactions between monomers dominate, a single polymer chain is presumed to
collapse  into a space-filling conformation. 
In a good solvent environment or at sufficiently high 
temperatures, a single polymer chain tends to
swell until its geometric structure bears similarity to 
a self-avoiding random walk (SARW). 
The collapsed phase  and the SARW phase are separated by $\Theta$-point  where a
polymer has the characteristics  of an 
ordinary random walk (RW). 
Biologically active proteins are commonly presumed to reside in the space filling collapsed phase,
under physiological conditions.  We shall examine proteins as an important subset
of polymers, for which a large amount of experimental data is available in 
PDB \cite{pdb}.

The phase where a polymer resides can be determined 
from the value of the scaling exponent  $\nu$ \cite{degennes,schafer,naka}.
To define this quantity, we consider the asymptotic behavior of  
the radius of gyration when the number of monomers $N$ is very large. With $ {\bf r}_i $ 
the coordinates of the skeletal atoms of the polymer,  the radius of gyration becomes in this limit 
\cite{degennes,schafer,naka,degen2,desc,sokal}
\begin{equation}
R_g \ = \ \sqrt{ \, \frac{1}{2N^2}  \sum_{i,j} ( {\bf r}_i  - {\bf r}_j )^2\, }   
\ \buildrel{N \ {\rm large}}\over{\longrightarrow}  \
R_0 N^{\nu}  + \dots
\la{R0}
\end{equation}
The length scale $R_0$ is an effective Kuhn distance between the
skeletal atoms  in the large-$N$ limit. It is  in principle  a computable quantity, 
that depends on all the atomic level details of the polymer and 
all the effects of environment including  pressure, temperature 
and chemical microstructure of the solvent. 
Unlike $R_0$,  the dimensionless scaling exponent  $\nu$ that governs  the large-$N$ 
asymptotic form of equation (\ref{R0}), 
is a universal quantity. Its numerical value is independent of the local atomic level structure,
and coincides with the inverse of the Hausdorff (fractal)  dimension $d_H$ of the polymer.
The numerical value of $\nu = 1/d_H$ serves as an order parameter of polymer phase structure:

For a continuous self-nonintersecting chain in 
three space dimensions $d_H$  can {\it in principle }
acquire any value between 1 and 3.  Simple examples of fractal structures where $d_H$ 
is not an integer,  include the piecewise linear Koch curve and attractors of chaotic equations
such as the Lorenz and the R\"ossler equations \cite{chaos}.  

The phases of polymers and in particular proteins, have been presumed to display 
fractal geometry \cite{frac0,frac1,dewey,frac2,frac3,dima,frac4,hong,huang,jcp,scheraga}.
Traditionally, the following values of $\nu$ are assigned to these phases \cite{degennes,schafer}: 
Biologically active proteins are commonly in the space filling $d_H = 3$ phase. 
For the $\Theta$-point $d_H = 2$, and for the SARW phase 
the Flory-Huggins value $d_H =  5/3$ is found \cite{huggins}, \cite{flory}.

. 

In analyses of PDB proteins,
additional values of $\nu$ have been proposed \cite{dewey}-\cite{scheraga}. For example,  \cite{hong} argues for the 
scaling exponent $\nu = 2/5$ instead of $1/3$ for the collapsed phase. Moreover, according to  
\cite{dewey}, \cite{hong} the value of $\nu$ depends on the type of the protein, different values 
are quoted for different fold types such as all-$\alpha$ proteins, all-$\beta$ proteins,
and $\alpha/\beta$ proteins. 
Furthermore, \cite{huang} argues that protein folding involves three stages:  
The Flory-Huggins value $\nu = 3/5$ proceeds to an  intermediate phase with $\nu = 3/7$, followed by  
$\nu = 2/5$ in the collapsed state. According to \cite{jcp} for unstructured proteins 
the scaling exponent has the value $\nu = 0.43 \pm 0.02$.

In this Letter  we introduce a novel geometric characteristic of
polymer phase structure, that we
call the folding angle. It is complementary to the scaling index $\nu$ and might
provide certain advantages: Unlike $\nu$ it can,
 in principle, be computed from a single structure. 
Moreover, (\ref{R0}) is known to be subject to very strong finite-size effects;   
in the SARW phase \cite{sokal} 
detects corrections to $\nu$ whenever $N$ is less than
$ N \sim 10^4$. Consequently, in the case of proteins where $N$ is much smaller,
the presence of potentially strong corrections to scaling effects in the value of $\nu$ should not be ignored.
The relation between $\nu$ and the folding angle could be a useful tool  to try and 
estimate these corrections. 

We consider a polymer backbone with 
skeletal atoms at $\mathbf r_i$. We define the unit length tangent ($\mathbf t$)
and bi-normal ($\mathbf b$) vectors at each site $i=1,...,N$,
\begin{equation}
\mathbf t_i \ = \ \frac{\mathbf r_{i+1} - \mathbf r_i}{|\mathbf r_{i+1} - \mathbf r_i|} \ \ \ \ \& \ \ \ \ \mathbf b_i \ = \
\frac{ \mathbf t_{i-1} \times \mathbf t_i}{| \mathbf t_{i-1} \times \mathbf t_i |}
\la{t}
\end{equation}
The backbone bond  ($\kappa$) and torsion  ($\tau$) angles are
\begin{equation}
 \mathbf t_{i+1} \cdot \mathbf t_i \ = \ \cos \kappa_i  \ \ \ \ \ \& \ \ \ \ \  
 \mathbf b_{i+1} \cdot \mathbf b_i  \ = \ \cos \tau_i
\la{kt}
\end{equation}
We assume that $N$ is very large. We 
introduce a block-spin transformation which at each step number $p$ combines two (or more) 
skeletal subunits into a new  subunit; see Figure 1.  
We introduce the new tangent vectors, corresponding to the new subunits, by setting
\[
\mathbf t^{(p)}_i  \ \to \ \frac{ \mathbf t^{(p)}_{2j-1} + \mathbf t^{(p)}_{2j}} { |\mathbf t^{(p)}_{2j-1} + 
\mathbf t^{(p)}_{2j}| } \ = \ \mathbf t_j^{(p+1)}
\]
\begin{figure}[h]
        \centering
                \includegraphics[width=0.4\textwidth]{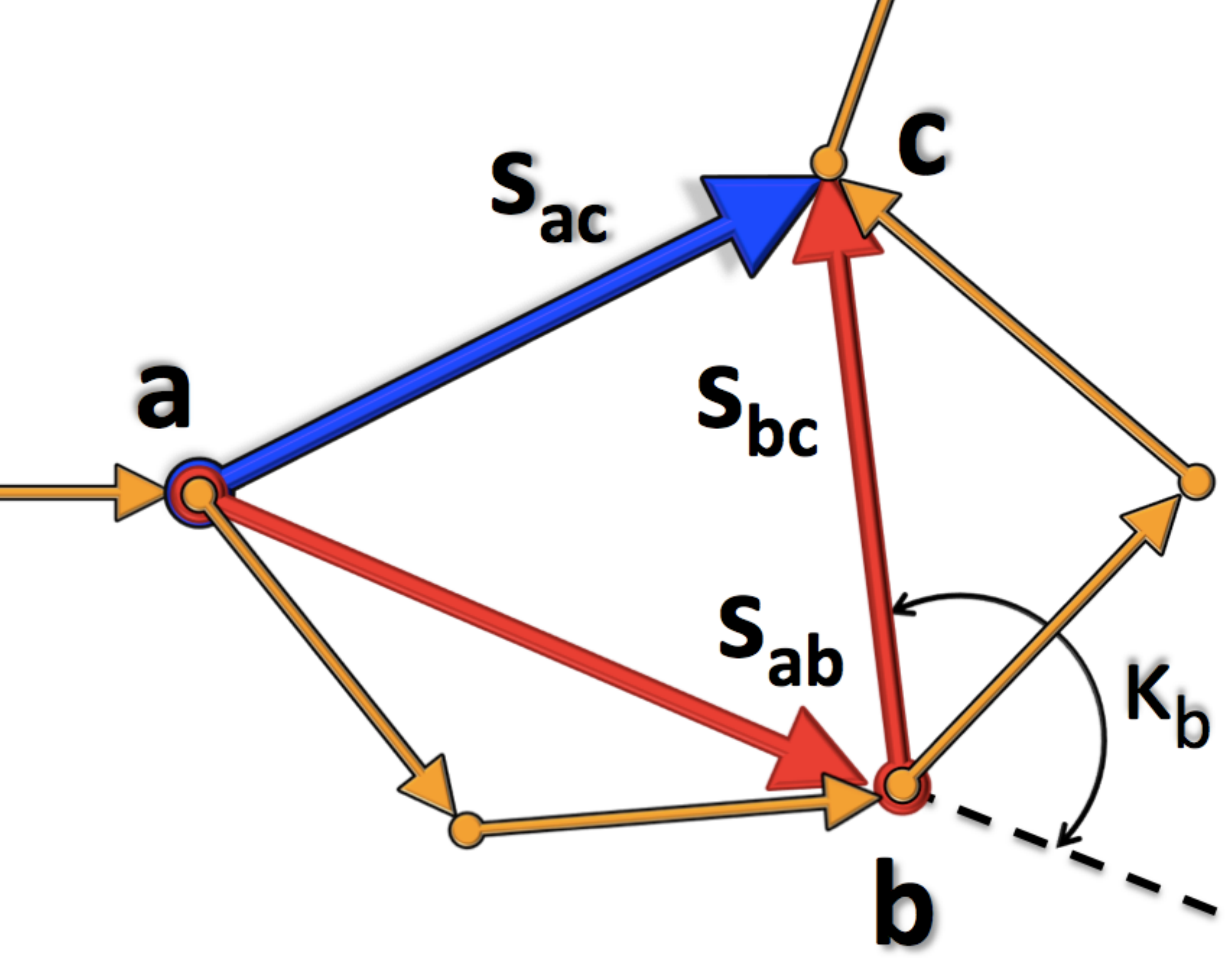}
        \caption{{ \it (Color online)
     Two steps in the block-spin transformation  that we utilize in evaluating (\ref{scale}).}}
       \label{Figure 2}
\end{figure}
This gives new coarse-grained bond angles,
\[
\cos\, \kappa_i^{{(p)}} = \mathbf t^{(p)}_{i+1} \cdot \mathbf t^{(p)}_{i}  
\to  \,  \mathbf t_{j+1}^{{(p+1)}} \cdot \mathbf t_{j}^{{(p+1)}}
\! =  \cos\, \kappa_j^{{(p+1)}}
\]
We assume that when we repeat this  transformation a very large number $p$ of times,
the numerical values of the transformed bond angles $\kappa^{(p)}_{i}$ converge 
towards a single fixed point value $\kappa^\star$,
\begin{equation}
< \! \mathbf t^{(p)}_i \cdot \mathbf t^{(p)}_{i+1} \! >  \
\buildrel{p \to \infty}\over{\longrightarrow}  \ \cos \kappa^\star
\la{scale}
\end{equation}
This is commonly the case for self-similar chains. 
We call $\kappa^\star$  the {\it folding angle} and propose 
that the numerical value of (\ref{scale}) depends only on the phase of polymer. 

We first argue that when the limit (\ref{scale}) exists and is unique,  there is the following asymptotic $p \to \infty$ relation 
between the cosine of $\kappa^\star$ and  the 
scaling exponent  $\nu$,
\begin{equation}
 \cos \kappa^\star  \approx  \ 2^{2\nu-1} -1 
\la{cval}
\end{equation}
\begin{equation}
 \Rightarrow \ \cos \kappa^\star  \ = \   
\left\{   \, \begin{matrix} -0.21  \ \dots  &   \hskip -2.2cm \, \nu = 1/3  \\
    \hskip -.8cm 0
&  \hskip -2.1cm  \nu = 1/2
\\ \hskip 0.2cm  0.11   \ \dots  & \hskip 0.cm   \nu = 0.5888 \dots \ \sim 3/5 \end{matrix} \right. 
\la{cval2}
\end{equation}
The SARW estimate is taken from  \cite{sokal} and 
$3/5$ is the corresponding Flory-Huggins 
result \cite{huggins}, \cite{flory}.  Note that the $\Theta$-point value $\nu = 1/2$ is 
exact: At the $\Theta$-point where
long range correlations along the chain are absent, standard arguments
\cite{degennes} imply that (\ref{scale}) must vanish.  The value for $\nu = 1/3$ is similarly exact,
for a space filling structure.

To justify (\ref{cval}), (\ref{cval2})  we consider a polymer chain where we have implemented the
block-spin transformation 
several times, to arrive at
a configuration where three consecutive nearest neighbor 
skeletal subunits that we denote by $a,b,c$ are connected by vectors $\mathbf s_{ab}
$ and $\mathbf s_{bc}$ as shown in Figure 1.  We introduce the next block-spin transformation.
As shown in Figure 1, it replaces these two vectors with the vector $\mathbf s_{ac}$.
We consider a statistical ensemble of the polymer, and compute the
ensemble average of the 
squared length 
of the vector $\mathbf s_{ac}$. The result is
\[
<\! \mathbf s_{ac}^2\! > \, = \, 
<\!\mathbf s_{ab}^2 \!> + <\!\mathbf s_{bc}^2\!> + 2<\!|\mathbf s_{ab}| \, |\mathbf s_{bc}|   \cos \kappa_{b}\! > 
\]
Consider the scaling limit where the block-spin subunits consist of  $n$ skeletal atoms, where $n$ is large. 
Assume that the polymer is in a phase where the distances $s_{ab}$, $s_{bc}$ and $s_{ac}$ 
scale according to (\ref{R0}), that is 
\[
<\! \mathbf s_{ab}^2\!> \, \sim \, <\! \mathbf s_{bc}^2\!> \, \sim \, n^{2\nu}  \ \ \ \&\ \ \ <\! \mathbf s_{ac}^2\!> \, \sim \, (2n)^{2\nu}
\]
Note that $\mathbf s_{ac}$ corresponds to the  step
where the skeletal subunit consists of $2n$ original
atomic  skeletals while both $\mathbf s_{ab}$  and $\mathbf s_{bc}$ are constructed with $n$ atomic skeletals.
We  now {\it assume}  that to leading order in $n$ the ensemble averages
factorize, so that we have
\[
<\! |\mathbf s_{ab}|\, | \mathbf s_{bc}| \cos \kappa_{b}\! > \, \sim \,  <\!|\mathbf s_{ab}|\!> <\! |\mathbf s_{bc}|\!>
 <\!\cos\kappa_{b}\!>  + \mathcal O(\frac{1}{n})
\]
From  this we immediately obtain (\ref{scale}) and (\ref{cval}), when
we identify  $\mathbf t_i^{ren}$ with the unit
vector in the direction $\mathbf s_{ab}$.  

Note  that (\ref{cval}) engages an exponential in
$\nu$.  Thus $\cos \kappa^\star$ might indeed be more sensitive than
$\nu$,  in characterizing corrections to scaling.

We first estimate $\cos \kappa^\star$,  in the case of
PDB  protein structures \cite{pdb}. Here space allows us to
analyze  in detail only the subset of mainly $\alpha$-helical
proteins in the CATH classification \cite{cath}; the issues raised in \cite{dewey}-\cite{jcp} will
be addressed  elsewhere. We consider those $\alpha$-helical proteins in PDB with less  than $30\%$ homology identity,
and single chain in biological assembly.
There are a total of 1174 structures in our data set, and this enables us to reliably
extend our analysis to $2n = 330$ residues.  The result is shown in Figure 2. We find that when $2n \sim  330 $ 
 \begin{equation}
<\! \cos \kappa^\star \! > \approx - 0.18 \dots  
\label{cosa1}
\end{equation}
which corresponds to $\nu \approx 0.357 \dots \ $ according to (\ref{cval}). This is remarkably  close to the value $\nu = 
1/3$ in (\ref{cval2}), 
for a fully space filling configuration.  
For comparison,  using the radius of gyration fit to the all-$\alpha$
PDB structures, \cite{hong} finds  $\nu \approx 0.403$ while \cite{scheraga} reports 
$\nu \approx 0.37$ for these structures. 

We observe in Figure 2 that when $n$ is very small,
$\cos \kappa^\star$  tends to have (mainly) positive values. Over a 
very short distance of only a few amino acids, the 
structure is determined by the covalent bonds. Indeed, due to steric constraints, 
the virtual C$_\alpha$ backbone bond angle is known to prefer values that are less than $\pi/2$ \cite{marty1,marty2}.
When the number of residues $n$ increases the backbone starts 
pulling together, (\ref{scale}) decreases and
becomes negative \cite{foot}.
When $n$ increases further, excluded volume effects come into play.
This  causes a dense-backing repulsion,  
the value of (\ref{scale}) starts increasing, converging 
towards the asymptotic value (\ref{cosa1}).

\begin{figure}[h]
        \centering
                \includegraphics[width=0.45\textwidth]{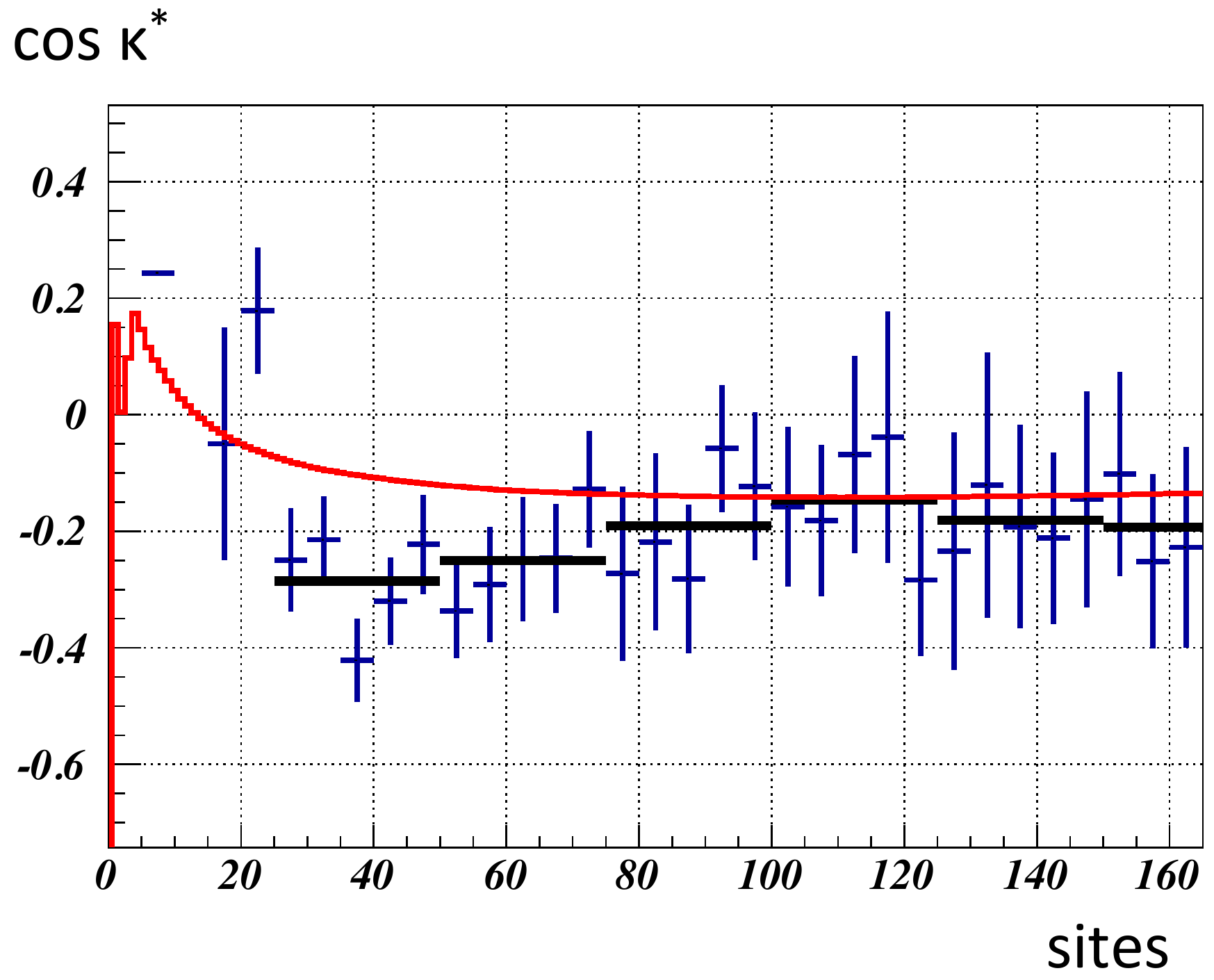}
        \caption{{ \it (Color online) 
      (Blue) entries denote distribution (\ref{scale}) in our PDB data,  described in the text. 
     The averages are over $\Delta N = 5$ bins. (Black) 
horizontal lines are piecewise linear interpolations that average the PDB data, over $\Delta N = 25$ 
bins, weighted over the number of PDB entries. Continuous (red) line is the 
result of theoretical computation using (\ref{E1}).
  }}
       \label{Figure 2}
\end{figure}

For a theoretical estimate of (\ref{scale}), we need a dynamical model. 
We have chosen the following Hamiltonian free energy \cite{ulf}-\cite{martin}.
\begin{equation}
E = \sum\limits_{i=1}^{N}  \biggl\{ - 2   \kappa_{i+1} \kappa_i   + 2 \kappa_i^2  + c
( \kappa_i^2 - m^2)^2 + b \kappa_i^2 \tau_i^2 
\biggr\} 
\ + \ \sum\limits_{i=1}^N 
\biggl\{  d \tau_i +  e  \tau_i^2
\biggr\} \ \ \ \ \ \ \ (\kappa_{N+1}=0) 
\la{E1}
\end{equation}
It describes collapsed 
chiral homopolymers as local energy minima
in terms of the backbone bond and torsion angles. 
The detailed derivation of (\ref{E1}) can be found in \cite{ying}. Here it suffices to  
state that the energy function (\ref{E1}) can be shown to be a long-distance limit that describes 
the full microscopic energy of a folded protein \cite{scheraga}.  
As such, it does not explain the details of the (sub-)atomic level mechanisms that give rise to protein folding.
In applications to polymers we need to complement (\ref{E1}) by  the excluded volume
constraint, due to steric repulsions.
We demand that the chain we construct using 
the ($\kappa_i, \tau_i$)
values in (\ref{E1}) by  inverting (\ref{kt}) and (\ref{t}),  is subject to 
\begin{equation}
| \mathbf r_i - \mathbf r_j |> \Delta \ \ \ \ \ i \not= j
\la{star}
\end{equation}
In the case of proteins we choose $\Delta = 3.8 ~ {\rm \AA}$, the  average distance between two neighboring $C_\alpha$
atoms. 

To compute the result shown by the red curve in Figure 2 we introduce a finite temperature environment
using  a canonical ensemble, 
and evaluate the ensuing Bolzmannian partition function numerically by Monte Carlo integration. 
We have collected statistics over
a period of around three months of wall-clock time, using a 120 processor MacPro desktop farm;  the error-bars are minuscule
and thus not displayed in Figure 2. In our simulations we thermalize
each chain during 10 million Monte Carlo steps with the following parameter values,
$ c= 5.4$,  $m = 1.51$,  $ b = 0.02$,  $d = -0.09$, $e =
-0.001$. 
These parameters are chosen so that the minimum energy configurations are like
$\alpha$-helical protein structures
\cite{nora,andrei1,martin}; we have checked that our results are quite insensitive to the choice of 
parameter values, and do not change if the number of Monte Carlo steps is increased.

From Figure 2 we observe that qualitatively, our numerical results  and the 
experimental PDB data are {\it quite } similar: For very small values on $n$, 
 (\ref{scale}) is positive. It then starts decreasing and becomes
negative. There is a minimum value, at $n_{min} \approx 110$.  
After this  (\ref{scale}) starts increasing towards
its asymptotic negative value of the scaling limit. For long chains $2n\sim 320$ we find
\begin{equation}
<\cos \kappa^\star > \ = \  - 0.14 ... 
\label{cosa2}
\end{equation}
corresponding to $\nu \approx 0.391 \dots$ when we use (\ref{cval}).  This is between
the values $\nu \approx 0.403$ and $\nu \approx 0.37$ reported  in \cite{hong,scheraga} respectively, for all-$\alpha$ proteins in PDB,
obtained by using  (\ref{R0}).

We have also estimated (\ref{cval}) from (\ref{E1}) in
the self-avoiding random walk phase, using an ensemble of  chains with {\it fixed} length $2n=500$. 
In the high temperature limit the energy does not contribute, only (\ref{star}) is relevant, and we find
\[
<\!\cos \kappa^\star \!> \ \approx   \, + \, 0.10 ...  
\]
From  (\ref{cval}) we now get $\nu \approx 0.57 $ which  is very close to the SARW value $\sim 0.5888 \ \dots $ 
obtained in \cite{sokal} for chains with $n \sim 10^5$. 

In summary, we have introduced the concept of folding angle as a new tool to characterize 
the phases of polymers.  We have proposed a relation 
between the cosine of the folding angle and the scaling exponent of the radius of gyration, that
we have investigated using both experimental data 
and model dependent numerical simulations. Unlike the scaling exponent the cosine of the 
folding angle can, in principle, be computed
from a single configuration.  The results also propose that the cosine of the folding angle 
could better reveal the presence 
of corrections to scaling in the vicinity of the collapsed state fixed point, 
than the scaling 
exponent. Thus we expect that the folding angle can become a valuable new order parameter
in understanding the phase structure of polymers.  
 
We acknowledge support from CNRS PEPS Grant, Region Centre Rech\-erche d$^{\prime}$Initiative Academique grant,
Sino-French Cai Yuanpei Exchange Program, and Qian Ren Grant at BIT.



\begin{thebibliography}{99}
\bibitem{naka2} Additional criteria include the density of vibrational states, 
and the probability of a random walker to be found at the point of origin as a function of time \cite{naka}.
\end{thebibliography}
\end{document}